\documentclass{iopart}
\usepackage{iopams}
\usepackage{graphicx,epsf}
\usepackage{subfig}
\usepackage{color}
\usepackage{mathrsfs}
\usepackage{dcolumn}
\usepackage{cite}
\usepackage{bm}

\begin{document}

\title[]{Core localized alpha-channeling via low frequency Alfv\'en  mode generation in reversed shear scenarios}

\author{Shizhao Wei$^1$, Tao Wang$^{1,2}$,  Liu Chen$^{1,3}$, Fulvio Zonca$^{2,1}$ and Zhiyong Qiu$^{1,2}$\footnote{E-mail: zqiu@zju.edu.cn}}

\address{$^1$Institute for    Fusion Theory and Simulation and Department of Physics, Zhejiang University, Hangzhou, P.R.C}
\address{$^2$ Center for Nonlinear Plasma Science and ENEA C. R. Frascati, Frascati, Italy}
\address{$^3$Department of   Physics and Astronomy,  University of California, Irvine CA 92697-4575, U.S.A.}

\begin{abstract}
A novel channel for fuel ions heating in tokamak core plasma is proposed and analyzed using nonlinear gyrokinetic theory. The channel is achieved via spontaneous decay of reversed shear Alfv\'en eigenmode (RSAE) into low frequency  Alfv\'en  modes (LFAM), which then heat  fuel ions via collisionless ion Landau damping.  The conditions for RSAE spontaneous decay are investigated, and the saturation level and the consequent fuel ion heating rate are also derived.   The channel is expected to be crucial for future reactors operating under reversed shear configurations, where fusion alpha particles are generated in the tokamak core where the magnetic shear is typically reversed, and there is a dense RSAE spectrum due to the small alpha particle characteristic dimensionless orbits.
\end{abstract}
\maketitle

\section{Introduction}
Energetic particles (EPs) including notably  fusion alpha particles related physics \cite{AFasoliNF2007,LChenRMP2016} are key elements  towards understanding  the performance of future fusion reactors such as ITER \cite{KTomabechiNF1991} and CFETR \cite{YWanNF2017}. Heating of  thermal ions via Coulomb collisions  is crucial for sustained fusion reactions, while EPs  excitation of collective oscillations  such as shear Alfv\'en wave (SAW) instabilities may lead to  EP loss and affect  the  confinement of thermal plasmas \cite{AFasoliNF2007,LChenRMP2016}. The SAW induced  EP anomalous transport   is determined by the saturation level and spectrum of SAWs  \cite{LChenJGR1999}.   Thus, it is crucial to   understand the dynamics of SAW instabilities that lead to their saturation \cite{FZoncaJPCS2021}. Another important topic in  EP research, is searching for alternative/complementary routes to transfer EP power to fuel ions,  i.e.,  alpha-channeling \cite{NFischPRL1992,NFischNF1994}, which is important in maintaining the self-sustained burning in future reactors, where collisional transfer of high energy fusion alpha is mostly due to electron  drag   due to the   high  alpha particle birth velocities.

Due to the magnetic geometry and plasma nonuniformities, the SAW continuous spectrum is characterized by various forbidden gaps, inside which different discrete SAW eigenmodes reside, e.g., toroidal Alfv\'en eigenmodes (TAE) due to toroidicity induced coupling of neighboring poloidal harmonics \cite{CZChengAP1985,LChenVarenna1988,GFuPoFB1989} and beta-induced Alfv\'en eigenmodes (BAEs) due to plasma compressibility \cite{WHeidbrinkPRL1993,FZoncaPPCF1996}.  Among various Alfv\'en eigenmodes (AEs), TAEs have drawn the most  attention in theoretical and numerical investigations, and are studied as  paradigm case for the  nonlinear dynamics of discrete AEs \cite{TSHahmPRL1995,LChenNF2007b,LChenRMP2016,HBerkPoFB1990c}, and the obtained general results can be applied to other SAW instabilities,  based on the   understanding of their respective linear properties. E.g.,  in future reactors operating in advanced reversed shear scenarios \cite{RAlbaneseFED2017,KTomabechiNF1991,YWanNF2017,JHuangNF2020} with minimized inductive current fraction, it is expected that reversed shear Alfv\'en eigenmodes (RSAEs, also known as Alfv\'en cascades) \cite{HBerkPRL2001,FZoncaPoP2002} are preferentially excited in tokamak center where fusion alpha particles are generated, while TAEs are excited in the relatively exterior region of the torus  with finite magnetic shear \cite{TWangPoP2018,ZRenNF2020}. On the other hand, BAEs, or more generally, low frequency Alfv\'en modes (LFAMs) in the frequency   range comparable or lower than BAEs \cite{FZoncaPPCF1996,FZoncaPoP1999,NGorelenkovPLA2007,IChavdarovskiPPCF2009,LChenPoP2017}, can be excited by both EPs as well as thermal plasmas due to their relatively low frequencies, in different toroidal mode numbers regimes. The properties of LFAMs in reversed shear plasmas,   including destabilization mechanism, mode polarization    dependence on  $q_{min}$,  are systematically investigated in Ref. \cite{RMaPPCF2022}. Here, $q_{min}$ is the local minimum of the safety factor $q$.

The linear properties of SAW instabilities expected in fusion reactors are extensively investigated \cite{LChenRMP2016,FZoncaPoP2014b,AFasoliNF2007,SSharapovNF2013}, and it is generally accepted that most unstable modes are characterized by   high-$n$ mode numbers with $k_{\perp}\rho_h\sim O(1)$ \cite{GFuPoFB1992,TWangPoP2018}, due to the competition of drive from EP pressure gradient   ($\propto n$)  and the stabilization by finite orbit width  effects (FoW) in the high-$n$ limit ($\propto 1/n$, noting the asymptotic form of Bessel functions accounting for FoW effects in the short wavelength limit  \cite{LChenPoP1994} \footnote{Note that, the FoW effects are formulated using Pad\'e approximation in Ref. \cite{LChenPoP1994}. For a more explicit expression, one may refer to equation (3) of  Ref. \cite{ZQiuPoP2016} for the EP response to TAE, assuming well circulating EPs.}).   Here, $n$ is the toroidal mode number, $k_{\perp}$ is the perpendicular wave number, and $\rho_h$ is the characteristic EP orbit width.  As a result, in future reactors such as ITER and CFETR \cite{KTomabechiNF1991,YWanNF2017} with $a/\rho_h\gtrsim O(10)$, the most unstable modes are characterized by $n\gtrsim O(10)$ with many modes having comparable linear growth rates \cite{TWangPoP2018}. Thus, nonlinear mode coupling is expected to be a channel for effectively saturating SAW instabilities  and modifying the perturbation spectrum \cite{FZoncaPRL1995,TSHahmPRL1995,LChenNF2001,LChenPRL2012,ZQiuPRL2018,ZQiuNF2019a}, among which, alpha-channeling through TAE decaying into modes prone to   ion Landau damping are proposed  and analyzed in Refs. \cite{TSHahmPRL1995,TSHahmPST2015,ZQiuPRL2018,JSeoNF2021}, which is shown to effectively transfer  fusion alpha particle power to fuel ions, in addition to nonlinearly saturate TAEs.

In this work, a new alpha-channeling mechanism, based on LFAM generation due to the nonlinear decay of RSAE,  is proposed and analyzed.    With the alpha particle characteristic orbit size much smaller than the tokamak minor radius \cite{AFasoliNF2007}, multiple RSAEs may be simultaneously excited \cite{LChenRMP2016,TWangPoP2018}, with their radial localization determined by $q_{min}$, and are thus, radially overlapped.  The RSAE   frequency is  determined by $\omega^2\simeq (nq_{min}-m)^2V^2_A/(q^2_{min}R^2_0)+\Delta^2_{\omega}$, with $\Delta_{\omega}$ being the deviation from local SAW continuum accumulation point due to $q$-curvature ($q''\equiv \partial^2 q/\partial r^2$) and non-perturbative EP drive \cite{TWangPoP2018}. Here, $m$ is  the  poloidal mode number, $V_A$ is the Alfv\'en speed, and $R_0$ is the major radius. Thus, for given $q_{min}$, RSAEs with different toroidal mode numbers  may have different parallel wave numbers $k_{\parallel} \equiv (nq_{min}-m)/(q_{min}R_0)$ and consequently frequencies covering TAE and BAE  frequency ranges, and two RSAEs may couple and generate  secondary modes with $\omega_{\pm}\simeq [(n_1\pm n_2)q_{min}-(m_1\pm m_2)]V_A/(q_{min}R_0)$. Among the two secondary modes, the lower frequency one, if satisfies the dispersion relation of a normal mode, e.g., BAE, or mor generally LFAM,     can be strongly driven unstable, and effectively heat thermal ions via LFAM Landau damping. This process, may provide a direct and efficient alpha-channeling mechanism  that transfers fusion alpha particle power to fuel ions.  This channel is of potential importance since RSAEs are expected to be firstly excited  in the tokamak center by  core localized alpha-particles \cite{TWangPoP2018},  and thus, the core localized power deposition   will heat core ions, leading to enhanced fusion performance.

Two independent processes can occur and lead to LFAM generation. In the first process, a linearly unstable RSAE spontaneously decays into another linearly stable RSAE and a low frequency sideband;  while in the second process, two linearly unstable RSAEs couple and generate a low frequency mode. The two processes can occur, since  there is a rich spectrum of (linearly stable or unstable) RSAEs and their kinetic counter-parts, i.e., kinetic RSAEs (KRSAE) \cite{FZoncaPoP1996,FZoncaPoP2014b,XWangPPCF2010} in reactors with $\rho_h\ll a$, so the frequency and wavenumber matching condition required for the resonant mode coupling process can be satisfied. In the present work, we will focus on the first process of parametric   instability of RSAE and discuss the condition for spontaneous decay; while the second process, which does not require an amplitude threshold condition, can be formally analyzed from the obtained nonlinear LFAM equation.

The rest of the manuscript is organized as follows. In Sec. \ref{sec:model}, the theoretical model of nonlinear gyrokinetic theory will be introduced. In Sec. \ref{sec:general_equation}, the linear particle responses to SAW instabilities in torus are reviewed, which are then used to  derive the general nonlinear equation describing the nonlinear interaction of SAW instabilities in torus. The nonlinear dispersion relation for RSAE parametric decay instability  is analyzed in  Sec. \ref{sec:NL_DR}. The consequences on RSAE saturation and core-localized ion heating is discussed in Sec. \ref{sec:saturation}. And finally, a brief summary and discussion are presented in Sec. \ref{sec:summary}.

\section{Theoretical model} \label{sec:model}

The governing equations describing nonlinear interactions among RSAEs and LFAM with all predominantly SAW polarization  can be derived from nonlinear gyrokinetic vorticity equation \cite{LChenJGR1991,LChenNF2001}
\begin{eqnarray}
&&\frac{c^2}{4\pi \omega^2_k}B\frac{\partial}{\partial l}\frac{k^2_{\perp}}{B}\frac{\partial}{\partial l}\delta \psi_k +\frac{e^2}{T_i}\left\langle (1-J^2_ k)F_0\right\rangle\delta\phi_k-\sum_{s=e,i}\left\langle\frac{q}{\omega_k}J_k\omega_d\delta H_k \right\rangle_s\nonumber\\
&=&-i\frac{1}{\omega_k}\sum_{\mathbf{k}=\mathbf{k}'+\mathbf{k}''}\Lambda^k_{k'',k'}\left [ \frac{c^2}{4\pi}k''^2_{\perp} \frac{\partial_l\delta\psi_{k'}\partial_l\delta\psi_{k''}}{\omega_{k'}\omega_{k''}} \right.\nonumber\\
&&\left.+ \left\langle e(J_kJ_{k'}-J_{k''})\delta L_{k'}\delta H_{k''}\right\rangle \right],
\label{eq:vorticityequation}
\end{eqnarray}
 and   quasi-neutrality condition
\begin{eqnarray}
\frac{n_0e^2}{T_i}\left(1+\frac{T_i}{T_e}\right)\delta\phi_k=\sum_{s=e,i} \left\langle q J_k\delta H_k \right\rangle_s,\label{eq:QN}
\end{eqnarray}
with the non-adiabatic particle response $\delta H_k$,   related to the perturbed particle  distribution function via $\delta f_k = -(q/T)F_0\delta\phi_k + \exp(-\mathbf{\rho}\cdot\nabla)\delta H_k$,  derived from nonlinear gyrokinetic equation \cite{EFriemanPoF1982}:
\begin{eqnarray}
\left(-i\omega+v_{\parallel}\partial_l+i\omega_d\right)\delta H_k=&-&i\omega_k\frac{q}{T}F_0J_k\delta L_k \nonumber\\
&-& \sum_{\mathbf{k}=\mathbf{k}'+\mathbf{k}''}\Lambda^k_{k'',k'} J_{k'}\delta L_{k'}\delta H_{k''}\label{eq:NLGKE}.
\end{eqnarray}

Here, the terms on the left hand side of  equation (\ref{eq:vorticityequation}) are, respectively, the field line bending, inertia and curvature coupling terms; while the terms on the right hand side are the formally nonlinear terms, corresponding to Maxwell and gyrokinetic Reynold stresses dominated by nonlinear electron and ion responses, respectively.  $\partial_l$ is the  derivative along the equilibrium magnetic field,  $J_k\equiv J_0(k_{\perp}\rho)$ with $J_0$ being the Bessel function of zero index accounting for finite-Larmor-radius effects (FLR), $\rho\equiv v_{\perp}/\Omega_c$, $\Omega_c=B_0q/(m c)$ is the cyclotron frequency,   $F_0$ is the equilibrium particle distribution function, and is taken as local Maxwellian for bulk electrons/ions, $\omega_d=(v^2_{\perp}+2 v^2_{\parallel})/(2 \Omega_c R_0) (k_r\sin\theta+k_{\theta}\cos\theta)$ is the magnetic drift frequency for a circular cross section large aspect ratio tokamak assumed in this work for simplicity of derivation. Furthermore,     $\Lambda^k_{k'',k'}\equiv (c/B_0)\hat{\mathbf{b}}\cdot\mathbf{k''}\times\mathbf{k'}$ with $\hat{\mathbf{b}}$ being the unit vector along the equilibrium magnetic field $\mathbf{B}_0$,   $\sum_{\mathbf{k}=\mathbf{k''}+\mathbf{k'}}$ denotes the selection rules of frequency and wavenumber matching conditions for nonlinear mode coupling,  $\delta L\equiv \delta\phi-k_{\parallel}v_{\parallel}\delta\psi/\omega$ with $\delta \psi\equiv \omega \delta A_{\parallel}/(c k_{\parallel})$ and $\delta A_{\parallel}$ being the parallel component of the vector potential, and ideal MHD condition is determined by $\delta\psi=\delta\phi$.

In this work, we assume the nonlinear coupling is dominated by thermal plasma contribution, while EPs, driving the pump RSAE unstable, contribute negligibly to the nonlinear coupling.  Consequently, the nonuniformity of thermal plasma in the tokamak core can be  neglected \cite{LChenPRL2021}, corresponding to thermal ion diamagnetic drift frequency  being smaller than BAE frequency here,   in consistency with the considered high-performance scenarios \cite{JHuangNF2020}. Thus, the present theory, which could be generalized to included kinetic ballooning modes (KBMs) \cite{FZoncaPPCF1996} and/or Alfv\'enic ion temperature gradient modes (AITGs) \cite{JKimPoFB1993,FZoncaPoP1999} by inclusion of thermal plasma nonuniformities,  is derived for application to BAE in its present form.   We consider  a spontaneous decay process, in which a pump RSAE decays into another linearly stable RSAE and a LFAM, and the condition for this process to occur is $\beta_i\ll1$ such that the frequency separation between RSAE and LFAM can easily be satisfied.  The nonlinear decay process, can be analyzed by deriving the nonlinear equations of the two sidebands, which can be coupled to yield the nonlinear parametric dispersion relation.   For proper evaluation of ion heating due to the LFAM Landau damping, the LFAM resonance with  thermal ions is crucial \cite{FZoncaPPCF1996}, and can be formally accounted for  in the anti-Hermitian part of the  LFAM dispersion relation.

\section{General nonlinear equation for resonant SAW three wave coupling}
\label{sec:general_equation}

The linear particle response to SAW instabilities can be derived by noting the $k_{\parallel}v_e\gg\omega\gg k_{\parallel}v_i\gtrsim \omega_d$ ordering, and one has, at the leading order, $\delta H_{e,k}\simeq -e F_0\delta\psi_k/T_e$ and $\delta H^{(0)}_{i,k}=e F_0J_k\delta\phi^{(0)}_k/T_i$,  which can be substituted into quasi-neutrality condition, and yield $\delta \phi^{(0)}_k=\delta \psi^{(0)}_k$, i.e., ideal MHD condition is maintained at the leading order.

To the next order, one derives
\begin{eqnarray}
\delta H^{(1)}_{i,k}=\frac{e}{T_i} F_0J_k\left[ \delta\phi^{(1)}_k+\frac{\omega_d}{\omega_k}\delta\phi^{(0)}_k \right], \nonumber
\end{eqnarray}
which can be substituted into quasi-neutrality condition and one obtains
\begin{eqnarray}
\delta\phi^{(1)}_k-\delta\psi^{(1)}_k=\frac{T_e}{T_i}\left\langle \frac{\omega_d}{\omega_k}\frac{F_0}{n_0}\right\rangle \delta\phi^{(0)}_k,\label{eq:e_parallel_compression}
\end{eqnarray}
i.e., breaking of ideal MHD constraint  due to plasma compressibility.   Finite parallel electric field generation due to FLR effects, i.e.,    kinetic Alfv\'en wave (KAW) related physics, will not be considered here for simplicity. However, as we show  later, the decay process favors higher-$n$ modes, for which inclusion of FLR may be needed  and can be accounted for straightforwardly \cite{LChenEPL2011}.

Substituting non-adiabatic particle responses  into vorticity equation, and noting   $\delta \phi^{(0)}_k=\delta \psi^{(0)}_k$, one derives the SAW mode equation in torus, i.e.,
\begin{eqnarray}
 \frac{n_0e^2}{T_i}b_k \mathcal{E}_k  \delta\phi^{(0)}_k=0, \label{eq:SAW_WKB}
\end{eqnarray}
with $\mathcal{E}_k\equiv   -k^2_{\parallel} V^2_A/\omega^2_k  + 1 - \omega^2_G/\omega^2_k $ being the SAW dielectric function in the WKB limit, and $\omega_G\equiv \sqrt{7/4+\tau} v_i/R_0$ being the leading order geodesic acoustic mode frequency \cite{NWinsorPoF1968,FZoncaEPL2008}, accounting for SAW continuum upshift and creation of beta-induced continuum gap.  We note that, in the expression of $\mathcal{E}_k$, effects of wave-particle interactions are not included, in consistency with the $k_{\parallel}v_i\ll\omega_k$ ordering for bulk non-resonant ions.  However, finite Landau damping due to resonance with ions  is crucial for alpha-channeling, and will be recovered formally in the later analysis by inclusion of the anti-Hermitian part of $\mathscr{E}_k$ \cite{FZoncaPPCF1996}.   Equation (\ref{eq:SAW_WKB}) is general, and can be applied to different modes in different scenarios. E.g., RSAE global dispersion relation can be derived by expanding $k^2_{\parallel}$ at $q_{min}$, and solving the eigenmode equation in Fourier-$k_r$ space \cite{FZoncaPoP2002}, while BAE physics  is dominated by $k^2_{\parallel}q^2R^2_0 \lesssim\beta_i$ \cite{FZoncaPPCF1996}.

Note that, for the present case of a pump RSAE decaying into another RSAE and a LFAM,  all  three modes involved   are  SAWs that satisfies $\omega^2\simeq k^2_{\parallel} V^2_A$ \footnote{Note that, LFAM dispersion relation can be quite different due to thermal plasma compression, but the general picture is the same,  especially for BAE of interest here with predominantly SAW polarization.}.    Thus, one can derive the general nonlinear equation for SAW nonlinear coupling, which can be applied to the case of RSAE nonlinear decay of interest.
Considering two SAWs, $\mathbf{\Omega}_1\equiv\Omega_1(\omega_1, \mathbf{k}_1)$ and $\mathbf{\Omega}_2\equiv\Omega_2(\omega_2, \mathbf{k}_2)$ coupling and generating a third mode, $\mathbf{\Omega}_3\equiv\Omega_3(\omega_3, \mathbf{k}_3)$, the nonlinear equations can be derived from nonlinear vorticity equation and quasi-neutrality condition. For simplicity of derivation, parallel force  balance equation is used instead of quasi-neutrality condition.

The first equation of $\mathbf{\Omega}_3$ mode  can be derived from nonlinear gyrokinetic vorticity equation,
\begin{eqnarray}
&& \frac{n_0e^2}{T_i} b_{k_3} \left(-\frac{k^2_{\parallel,3}V^2_A}{\omega^2_{3}} \delta\psi_{k_3} + \delta\phi_{k_3} -\frac{\omega^2_G}{\omega^2_{3}} \delta\phi_{k_3}\right)\nonumber\\
&\simeq& -\frac{i}{\omega_{3}} \Lambda^{k_3}_{k_2,k_1} \left[\frac{c^2}{4\pi}(k^2_{\perp,1}-k^2_{\perp,2})\frac{k_{\parallel,1}k_{\parallel,2}}{\omega_{1}\omega_{2}} \delta\psi_{k_1}\delta\psi_{k_2} \right.\nonumber\\
&&\left. + \langle e(J_{k_1}-J_{k_2})(\delta L_{k_1}\delta H_{i,k_2}+\delta L_{k_2}\delta H_{i,k_1})\rangle\right]. \label{eq:VT_k3}
\end{eqnarray}
In deriving equation (\ref{eq:VT_k3}), we have noted, in the nonlinear Reynolds stress \cite{FZoncaPoP2004,LChenPRL2012},  $\delta\phi_k\simeq\delta\psi_k$, $\delta H_{i,k}\simeq e F_0\delta\phi_k/T_i$, and neglected $O(k^2_{\perp}\rho^2_i)$ order corrections.  Substituting the lowest order ion response into the Reynolds stress term,   the nonlinear vorticity equation of $\mathbf{\Omega}_3$ then becomes
\begin{eqnarray}
&&b_{k_3}\left(-\frac{k^2_{\parallel,3}V^2_A}{\omega^2_{3}} \delta\psi_{k_3} + \delta\phi_{k_3} -\frac{\omega^2_G}{\omega^2_{3}} \delta\phi_{k_3}\right)\nonumber\\
&\simeq&-\frac{i}{\omega_{3}} \Lambda^{k_3}_{k_2,k_1} (b_{k_2}-b_{k_1}) \left(1-\frac{k_{\parallel,1}k_{\parallel,2} V^2_A}{\omega_{1}\omega_{2}}\right)\delta\phi_{k_1}\delta\phi_{k_2}.\label{eq:VT_k3_f}
\end{eqnarray}

The other equation can be derived from parallel electron force balance equation,   and  one has,
\begin{eqnarray}
\delta\phi_{k_3}-\delta\psi_{k_3} = -i \Lambda^{k_3}_{k_2,k_1}\frac{1}{k_{\parallel,3}}\left(\frac{k_{\parallel,1}}{\omega_{1}} - \frac{k_{\parallel,2}}{\omega_{2}}\right) \delta\phi_{k_1}\delta\phi_{k_2},\label{eq:QN_k3_f}
\end{eqnarray}
i.e., nonlinear extension of ideal MHD constraint, in addition to plasma compressibility as shown in equation (\ref{eq:e_parallel_compression}).

Substituting equation (\ref{eq:QN_k3_f}) into (\ref{eq:VT_k3_f}), one obtains
\begin{eqnarray}
b_{k_3}\mathcal{E}_{k_3}\delta\phi_{k_3} &=&- \frac{i}{\omega_3} \Lambda^{k_3}_{k_2,k_1} \left[  (b_{k_2}-b_{k_1})\left(1-\frac{k_{\parallel,1}k_{\parallel,2}V^2_A}{\omega_{1}\omega_{2}}\right) \right.\nonumber\\
&&\left.+b_{k_3}V^2_A\frac{k_{\parallel,3}}{\omega_3}\left(\frac{k_{\parallel,1}}{\omega_{1}} - \frac{k_{\parallel,2}}{\omega_{2}}\right)  \right]\delta\phi_{k_1}\delta\phi_{k_2}.\label{eq:NL_SAW}
\end{eqnarray}

Equation (\ref{eq:NL_SAW}) describes the nonlinear evolution of SAWs, as $\mathbf{\Omega}_3$ being modified by the beating of $\mathbf{\Omega}_1$ and $\mathbf{\Omega}_2$,  with the first term on the right hand side from the competition of Reynolds and Maxwell stresses and the second term from finite parallel electric field contribution to field line bending term. Note that,  $(\omega_1+\omega_2)\simeq (k_{\parallel,1}+k_{\parallel,2})V_A$,     $\mathbf{\Omega}_3$  naturally satisfies the SAW D.R., and can be strongly excited  if it is a normal mode of the system,  leading to  significant spectrum evolution of SAW turbulence.  Note that, though the primary motivation of the present work is investigating the LFAM generation by RSAEs, equation (\ref{eq:NL_SAW}) can be applied to study the nonlinear SAW couplings in the high frequency range, e.g., the nonlinear coupling among TAE,  ellipticity induced AE (EAE) and non-circular AE (NAE), whose frequency matching  condition can be naturally satisfied.   It can also be generalized to kinetic Alfv\'en waves (KAW) \cite{AHasegawaPoF1976} by  properly accounting for parallel electric fields due to kinetic effects \cite{LChenEPL2011},  which is expected to be crucial  due to the intrinsic nonuniformity in magnetically confined plasmas.     Equation (\ref{eq:NL_SAW}) can be more generally applied for KAW spectral cascading in space plasmas,  e.g., solar wind,  by neglecting the $\omega^2_G$ term that is unique in torus,   and $k_{\parallel}$  can be more flexibly taken  without the   periodicity constraint in a torus.

\section{Parametric decay of RSAE}
\label{sec:NL_DR}

Equation (\ref{eq:NL_SAW}) will be applied to the nonlinear decay of a pump RSAE   $\mathbf{\Omega}_0\equiv\Omega_0(\omega_0, \mathbf{k}_0)$  into a RSAE sideband  $\mathbf{\Omega}_1\equiv\Omega_1(\omega_1, \mathbf{k}_1)$ and a LFAM  $\mathbf{\Omega}_B\equiv\Omega_B(\omega_B, \mathbf{k}_B)$, with the frequency/wavenumber matching condition $\mathbf{\Omega}_0=\mathbf{\Omega}_1+\mathbf{\Omega}_B$ assumed without loss of generality.
For RSAE and LFAM being dominated by single-$n$ and single-$m$ mode structures, we take
\begin{eqnarray}
\delta\phi_k=A_k(t)\Phi_k(x) \exp{\left(-i\omega_k t+in\xi-im\theta\right)},
\end{eqnarray}
with  $A_k(t)$ being the slowly varying mode amplitude, $\Phi_k(x)$ the parallel mode structure localized about $q_{min}$ with $x\equiv nq-m$, and    the normalization condition $\int |\Phi_k|^2 dx=1$ is satisfied.
 For the effective transfer of alpha particle energy to core ions, $\omega_B\lesssim O(v_i/(qR_0))$, and thus, $|\omega_B|\ll |\omega_0|, |\omega_1|$ and $k_{\parallel,B}\simeq 0$. Thus, the $q_{min}$ surface where secondary $\mathbf{\Omega}_B$ locates, also corresponds to the rational surface of $\mathbf{\omega}_B$, i.e., $\mathbf{\Omega}_B$ is the LFAM in the reversed shear configuration, as investigated experimentally \cite{WHeidbrinkNF2021} and theoretically \cite{RMaPPCF2022}.  We then have, $\omega_0\simeq\omega_1$ and $k_{\parallel,0}\simeq k_{\parallel,1}$.  Effects of small frequency mismatch on the decay process will be discussed later.

The nonlinear RSAE  sideband equation can be derived  from equation (\ref{eq:NL_SAW}) as
\begin{eqnarray}
b_1\mathcal{E}_1\delta\phi_1&=& -\frac{i}{\omega_{1}} \Lambda^{k_1}_{k_0,k_{B^*}}\alpha_1 \delta\phi_{0}\delta\phi_{{B^*}}, \label{eq:RSAE_sideband}
\end{eqnarray}
with  $\alpha_1\equiv (b_0-b_B)(1- k_{\parallel,B}k_{\parallel,0}V^2_A/(\omega_0\omega_B))  +  b_1 V^2_A  (k_{\parallel,1}/\omega_{1} ) (k_{\parallel,B}/\omega_{B} - k_{\parallel,0}/\omega_{0})$.  The nonlinear $\Omega_1$ eigenmode equation can be derived from equation (\ref{eq:RSAE_sideband}), by multiplying both sides by $\Phi_0$, and averaging over radial mode structure, and one obtains
\begin{eqnarray}
\hat{b}_1\hat{\mathcal{E}}_1 A_1 =-\frac{i}{\omega_1} \left\langle \Lambda^{k_1}_{k_0,k_{B^*}}\alpha_1 \Phi_1\Phi_0\Phi_{B}\right\rangle_x A_0 A_{B^*}, \label{eq:RSAE_sideband_eigen}
\end{eqnarray}
with $\langle \cdots\rangle_x\equiv \int \cdots dx$ denoting   averaging over the fast radial scale, and $\hat{b}_1\hat{\mathcal{E}}_1\equiv \int \Phi_1 b_1 \mathcal{E}_1\Phi_1 dx$ being the $\Omega_1$ eigenmode dispersion relation.

The nonlinear LFAM equation, on the other hand, can be derived as
\begin{eqnarray}
b_B\mathcal{E}_B\delta\phi_{B}&=& -\frac{i}{\omega_{B}} \Lambda^{k_B}_{k_0,k_{1^*}} \alpha_B \delta\phi_{0}\delta\phi_{{1^*}}, \label{eq:LFAM}
\end{eqnarray}
with $\alpha_B \equiv (b_0-b_1)(1- k_{\parallel,1}k_{\parallel,0}V^2_A/(\omega_0\omega_1))  +  b_B V^2_A  (k_{\parallel,B}/\omega_{B} ) (k_{\parallel,1}/\omega_{1} - k_{\parallel,0}/\omega_{0})$.   The LFAM eigenmode equation can be derived similarly, and one obtains
\begin{eqnarray}
\hat{b}_B\hat{\mathcal{E}}_B A_B=-\frac{i}{\omega_B}  \left\langle \Lambda^{k_B}_{k_0,k_{1^*}}\alpha_B \Phi_B\Phi_0\Phi_{1}\right\rangle_x A_0 A_{1^*}, \label{eq:LFAM_eigen}
\end{eqnarray}
with $\hat{b}_B\hat{\mathcal{E}}_B$ being the LFAM eigenmode dispersion relation.
Equations (\ref{eq:RSAE_sideband_eigen}) and (\ref{eq:LFAM_eigen}) are readily reduced to the  nonlinear eigenmode equations of RSAE sideband and LFAM   in the WKB limit, and can be simplified noting the respective parameter regimes.

The parametric decay dispersion relation for RSAE nonlinear decaying into another RSAE and  LFAM, can then be derived, by combining equations (\ref{eq:RSAE_sideband_eigen}) and (\ref{eq:LFAM_eigen})
\begin{eqnarray}
\hat{\mathcal{E}}_1\hat{\mathcal{E}}_{B^*} \simeq  \left(\hat{\Lambda}^{k_1}_{k_0,k_{B^*}}\right)^2\frac{\hat{\alpha}_N}{\hat{b}_B\hat{b}_1 \omega_{B}\omega_{1}} \hat{C}^2 |A_0|^2, \label{eq:parametric_disp}
\end{eqnarray}
with  $\hat{C}\equiv \langle\Phi_{0}\Phi_B\Phi_1\rangle_x$,  $\hat{\alpha}_N\equiv \hat{\alpha}_1\hat{\alpha}_B$ with $\hat{\alpha}_k=\langle \alpha_k\rangle_x$,  $ \hat{\Lambda}^{k_1}_{k_0,k_{B^*}}=  \langle\Lambda^{k_1}_{k_0,k_{B^*}}\rangle_x$, and
$ \Lambda^{k_1}_{k_0,k_{B^*}} =   \Lambda^{k_B}_{k_0,k_{1^*}}$ noted. In deriving equation (\ref{eq:parametric_disp}), we have noted that  $\hat{\alpha}_N$  and
$ \left(\hat{\Lambda}^{k_1}_{k_0,k_{B^*}} \right)^2$   are operators acting on the respective mode structures, and they are moved out of the spatial averaging in that they are both predominantly even operators with respect to $q_{min}$ surface.  The integration $\langle\Phi_{0}\Phi_B\Phi_1\rangle_x$ can be evaluated, noting that $\Phi_B$ typically has a much narrower structure than those of $\Phi_0$ and $\Phi_1$. Taking $\Phi_k\simeq \exp(-x^2/2\Delta^2_k)/(\pi^{1/4}\Delta^{1/2}_k)$ with $\Delta_k$ being the characteristic radial width of the parallel mode structure, we have,   $\hat{C}\simeq \sqrt{2 \Delta_B/(\sqrt{\pi}\Delta_0\Delta_1)}$, with $\Delta_0\sim\Delta_1\sim O(1)$ and $\Delta_B\sim O(\beta^{1/2})$.

Expanding $\hat{\mathcal{E}}_{1}\simeq i \partial_{\omega_1}\hat{\mathcal{E}}_{1}(\partial_t+\gamma_1)\simeq (2 i/\omega_1) (\gamma+\gamma_1)$ and $\hat{\mathcal{E}}_{B^*}\simeq (-2i/\omega_B) (\gamma+\gamma_B)$, with $\partial_{\omega_1}\hat{\mathcal{E}}_{1}\equiv \partial \hat{\mathcal{E}}_{1}/\partial \omega_1$, $\gamma$ denoting the slow temporal variation of $\mathbf{\Omega}_1$ and $\mathbf{\Omega}_B$ due to the parametric instability, and $\gamma_1/\gamma_B$ being the linear damping  rates of RSAE/LFAM imbedded in the anti-Hermitian part of $\mathcal{E}_1/\mathcal{E}_B$,  one obtains
\begin{eqnarray}
(\gamma+\gamma_1)(\gamma+\gamma_B)=\left(\hat{\Lambda}^{k_1}_{k_0,k_{B^*}}\right)^2 \frac{\hat{\alpha}_N}{4 \hat{b}_B \hat{b}_1}  \hat{C}^2|A_0|^2. \label{eq:parametric_DR}
\end{eqnarray}
The condition for the pump RSAE spontaneous decay can thus be obtained from equation (\ref{eq:parametric_DR}) as
\begin{eqnarray}
\hat{\alpha}_N>0,
\end{eqnarray}
    and
\begin{eqnarray}
\frac{(\hat{\Lambda}^{k_1}_{k_0,k_{B^*}})^2}{4 \hat{b}_B \hat{b}_1}\hat{\alpha}_N  \hat{C}^2|A_0|^2 >\gamma_B\gamma_1 \label{eq:threshold}
\end{eqnarray}
for the nonlinear drive overcoming the threshold due to $\mathbf{\Omega}_1$ and $\mathbf{\Omega}_B$ Landau damping.

The nonlinear dispersion relation is very complex, and depends on various conditions including the polarization   and mode structure  of the three modes involved.  For the analytical progress,    the WKB limit  and  the strong assumption of $k_{\parallel,B}\rightarrow 0$ is adopted   from now on, and we have $k_{\parallel,0}\simeq k_{\parallel, 1}$ and thus, $\hat{\alpha}_N$ can be simplified as
\begin{eqnarray}
\hat{\alpha}_N\simeq \left(b_0-b_1\right) \left(1-\frac{k_{\parallel,1}k_{\parallel,0}V^2_A}{\omega_0\omega_1}\right)   \left(b_0-b_B- b_1 \right).
\end{eqnarray}

The sign of $\hat{\alpha}_N$ is determined by several factors. First, the sign of  $1-k_{\parallel,1}k_{\parallel,0}V^2_A/(\omega_0\omega_1)$ is determined by $q_{min}$ and $n_0/n_1$ that determine the respective SAW continuum structure, i.e., whether the RSAEs are localized below the local minimum of SAW continuum or above the local maximum; $b_0-b_1$ is determined by the respective toroidal mode numbers $n_0/n_1$, noting $k_{\theta}\propto n q/r$ and $k_r\propto \sqrt{q'' n^2/q_{min}}$.   Noting that, $\mathbf{k}_{\perp,0}=\mathbf{k}_{\perp,B}+\mathbf{k}_{\perp,1}$, we have, $b_0-b_B-b_1= (\mathbf{k}_{\perp,0}\cdot\mathbf{k}_{\perp,1}-k^2_{\perp,1}) \rho^2_i$, and its sign is positive for $\cos\eta>|k_{\perp,1}/k_{\perp,0}|$, with $\eta$ being the angle between $\mathbf{k}_{\perp,0}$ and $\mathbf{k}_{\perp,1}$.    The above three conditions are quite complicated, but two parameter regimes can be identified for the spontaneous decay process to occur. The first parameter regime corresponds to $k_{\perp,1}\gg k_{\perp,0}$, such that $(b_0-b_1)(b_0-b_B-b_1)>0$; and $\hat{\alpha}_N>0$ can be satisfied with $1-k_{\parallel,0}k_{\parallel,1}V^2_A/(\omega_0\omega_1)>0$,  which generally requires  $\mathbf{\Omega}_1$ being excited above the local SAW continuum accumulation point with $n_1q_{min}< m_1$ \footnote{Note $n_1q_{min}< m_1$ is a sufficient condition for $1-k_{\parallel,0}k_{\parallel,1}V^2_A/(\omega_0\omega_1)>0$, but not necessary.}.

Another parameter regime can be found with $1-k_{\parallel,0}k_{\parallel,1}V^2_A/(\omega_0\omega_1)<0$, i.e., to have $\mathbf{\Omega}_1$ being excited below the local minimum of SAW continuum. In this case, $\hat{\alpha}_N>0$ can be satisfied with $(b_0-b_1)(b_0-b_B-b_1)<0$, which requires $b_1<b_0$ while $\mathbf{k}_{\perp,0}\cdot\mathbf{k}_{\perp,1} < k^2_{\perp,1}$.

We note  that, the $\Lambda^{k_1}_{k_0,k_{B^*}}$ in the nonlinear coupling cross-section  is maximized for $\mathbf{k}_{\perp,1}$ being perpendicular to $\mathbf{k}_{\perp,0}$,    and, also, that  the linearly unstable pump RSAE  typically have relatively broad mode structures   with $k_{\perp}\rho_h\sim O(1)$ \cite{TWangPoP2018}. For finite LFAM generation with the mode structures typically much narrower than that of the pump RSAE by $O(\omega_B/\omega_0)\sim O(\sqrt{\beta}_i)$,  the nonlinear coupling, thus, may occur in the radially fast varying  inertial  layer of $\Omega_1$. As a result, the toroidal mode number of $\Omega_1$ is expected to be much higher than that of the pump RSAE, which makes the first parameter region with $b_1\gg b_0$ optimized for the nonlinear process.  On the other hand,
the coefficient on the left hand side of equation (\ref{eq:threshold}) is proportional to $b_1$ considering $b_1\simeq b_B\gg b_0$, which further confirms the first parameter regime with $b_1\gg b_0$ is favored, i.e., normal cascading to high-$n_1$ regime. The upper bound of $n_1$ can be determined by $\gamma_1(n_1)$, i.e., the decay RSAE $\mathbf{\Omega}_1$ damping rate increases with $n_1$, as is shown in equation (\ref{eq:threshold}).
The condition for wavenumber/frequency matching condition to be satisfied, especially the requirement on $|\omega_B|$ being comparable with ion transit frequency, can be satisfied due to the potential dense RSAE/kRSAE spectrum in the high-$n$ limit.

The threshold condition for the RSAE spontaneous decay, for the first case of ``normal cascading",  can be estimated from equation (\ref{eq:threshold}), and one obtains
\begin{eqnarray}
\left|\frac{\delta B_{\perp,0}}{B_0}\right|^2 &>& \frac{4\gamma_1\gamma_B}{\omega_0\omega_1} \frac{k^2_{\parallel,0}}{k^2_{\perp,1}} \frac{1}{\hat{C}^2}  \frac{1}{1-k_{\parallel,0}k_{\parallel,1}V^2_A/(\omega_0\omega_1)}                  \nonumber\\
&\sim& \mathcal{O}(10^{-7}),
\end{eqnarray}
and is   comparable with or slightly higher than typical threshold condition for other dominant nonlinear mode coupling processes, e.g., ZS generation \cite{LChenPRL2012,SWeiJPP2021}. This threshold amplitude, is also  consistent with typical SAW instability intensity observed in experiments \cite{WHeidbrinkPRL2007}. Thus, this channel can be an important process in determining the nonlinear dynamics of RSAE, and the consequent transport by the short wavelength RSAE sideband and nonlinear thermal ion heating via the nonlinearly generated LFAM.
In deriving the threshold, typical parameter are used, i.e.,  $\gamma/\omega\sim 10^{-2}$,  $k_{\perp,1}\sim 1/\rho_i$, $k_{\parallel}\sim 1/R_0$,  $k_{\parallel}/k_{\perp,1}\sim \rho_i/R_0\sim 10^{-3}$,  $\Delta_B/\Delta_0\sim O(\omega_B/\omega_0)\sim 10^{-1}$,  and $|1-k_{\parallel,1}k_{\parallel,0}V^2_A/(\omega_0\omega_1)|\sim|\gamma/\omega_0| \sim \mathcal{O}(10^{-2})$ is assumed.

\section{Nonlinear saturation and core-localized ion heating}
\label{sec:saturation}

The RSAE saturation level   can be estimated by considering the feedback of the two sidebands to the pump RSAE, which can be derived from equation (\ref{eq:NL_SAW}) as
\begin{eqnarray}
\hat{b}_0\hat{\mathcal{E}}_0 A_0\simeq -\frac{i}{\omega_{0}}\hat{\Lambda}^{k_0}_{k_1,k_{B}} \hat{\alpha}_0 \hat{C} A_1 A_B,\label{eq:RSAE_pump}
\end{eqnarray}
with $\alpha_0= (b_1-b_B)  (1- k_{\parallel,B}k_{\parallel,1}V^2_A/(\omega_1\omega_B))  +  b_0 V^2_A(k_{\parallel,0}/\omega_{0})  (k_{\parallel,B}/\omega_{B}- k_{\parallel,1}/\omega_{1}) $.

Expanding equations (\ref{eq:RSAE_sideband_eigen}), (\ref{eq:LFAM_eigen}) and (\ref{eq:RSAE_pump}) along their characteristics, one obtains
\begin{eqnarray}
\left(\partial_t+\gamma_1\right) A_1&=&-\frac{\hat{\alpha}_1}{\hat{b}_1\omega_1 \partial_{\omega_1}\mathcal{E}_{1,\mathcal{R}}} \hat{\Lambda}^{k_1}_{k_0,k_{B^*}} \hat{C} A_0A_{B^*},\label{eq:3wave_1}\\
\left(\partial_t+\gamma_B\right) A_{B^*}&=& \frac {\hat{\alpha}_B}{\hat{b}_B\omega_B \partial_{\omega_{B^*}}\mathcal{E}_{B^*,\mathcal{R}}} \hat{\Lambda}^{k_B}_{k_0,k_{1^*}} \hat{C} A_{0^*}A_1,\label{eq:3wave_2}\\
\left(\partial_t-\gamma_0\right)A_0&=&-\frac{\hat{\alpha}_0}{\hat{b}_0\omega_0\partial_{\omega_0}\mathcal{E}_{0,\mathcal{R}}} \hat{\Lambda}^{k_0}_{k_1,k_B} \hat{C} A_1A_B,\label{eq:3wave_3}
\end{eqnarray}
and the saturation level of pump RSAE and LFAM can be evaluated from the coupled three wave equations. We note that, the  coupled nonlinear equations with drive and dissipation can exhibit different dynamics from limited cycle oscillation to period doubling and finally route to chaos,  depending on the driving/dissipation as well as the initial conditions \cite{GWeiAPS2021}. Here, an order of magnitude estimation of the  saturation level  can be  derived from the fixed point solution, among which the LFAM saturation level can be derived from equations (\ref{eq:3wave_1}) and (\ref{eq:3wave_3}). One obtains,
$|A_B|^2=  \gamma_0\gamma_1 \hat{b}_0\hat{b}_1\omega_0\omega_1\partial_{\omega_1}\mathcal{E}_{1,\mathcal{R}}\partial_{\omega_0}\mathcal{E}_{0,\mathcal{R}}/(\hat{\alpha}_0 \hat{\alpha}_1 |\hat{C}|^2 (\hat{\Lambda}^{k_0}_{k_1,k_B})^2)$, and the ion heating rate due to LFAM Landau damping, can be estimated as
\begin{eqnarray}
P_i=2\gamma_B \omega_B\frac{\partial \mathscr{E}_{B,\mathcal{R}}}{\partial\omega_B} |A_B|^2.
\end{eqnarray}
The obtained core ion heating due to LFAM conllisionless damping, is expected to be complementary to Coulomb collisional heating, which is less effective in the tokamak center. This channel, achieved via the Landau damping of secondary LFAM, noting that $k_{\parallel,B} \ll1$, is highly localized around the $q_{min}$ surface (this conclusion can also be obtained, noting  as the ``secondary" LFAM structure will be determined by the primary RSAE, with a narrower extent than the primary RSAEs), will deposit fusion alpha particle power locally and heating core ions, leading to direct improvement of fusion performance in the tokamak center.

\section{Conclusion and Discussion}
\label{sec:summary}

In conclusion, a novel channel for RSAE nonlinear saturation   is proposed and analyzed, which  is expected to be important in regulating SAW instability induced alpha particle transport and heating of   fuel ions   in  future  reactors burning plasmas.  The saturation  is achieved through the spontaneous decay of the unstable pump RSAE into  linearly stable  RSAE decay wave and LFAM,  while the fuel ion heating is achieved through the ion Landau damping of the secondary LFAM. The conditions for the RSAE spontaneous decay is analyzed. It is found that decay into   RSAE sideband with higher toroidal mode number is preferred, and the threshold condition on pump RSAE amplitude is derived, which is compatible with typical experimentally observed SAW instability amplitudes. The saturation levels of RSAE and LFAM are estimated  from the fixed point solution of the coupled nonlinear equations, from which the fuel ion heating rate  is also derived.
This channel is expected to be relevant and crucially  important  for reactors since RSAEs are expected to be firstly excited by core localized fusion alpha particles in the high performance advanced reversed shear scenarios,  and the resulting core localized fuel ion heating will directly contribute to the performance of the reactor.  An implication from the present analysis, is the potential importance of rational $q_{min}$ that may lead to low-order rational surfaces, which can yield broader mode structure and thus stronger couplings.

The present analysis  focused on the picture of RSAE nonlinear decay, while neglected the effects of thermal plasma nonuniformity. The derivation, also assumed SAW polarization of all the three modes involved. Thus, the present analysis, while can be directly applied to RSAE decay into BAE in the present form,  cannot be directly applied to RSAE decay into KBM or AITG, where plasma nonuniformity is crucial for the mode presence, and the mode may have a finite parallel electric field \cite{RMaPPCF2022}. The generalization of the present analysis, to include system nonuniformity, can be tedious but straightforward, and will be carried out in a separate work.

As a final remark, several channels may contribute to the RSAE nonlinear saturation, e.g., self-consistent re-distribution of EPs \cite{TWangPoP2019}, and  zonal field generation generation \cite{SWeiJPP2021} with notably the effects of zonal current on modifying the local SAW continuum structures. For the proper evaluation of EP confinement, the nonlinear dynamics  of RSAE including   saturation level is required, which  is determined by the relative importance of these channels; and  thus, more in-depth investigation including nonlinear gyrokinetic simulation \cite{ZLinScience1998} is required. In fact,  the effects investigated in the present work, e.g., the ion polarization nonlinearity unique to  gyrokinetic ion Reynold stress,   can only be captured by  simulations with gyrokinetic thermal ions.

\section*{Acknowlodgement}

This work is supported by the National Key Research and Development  Program of China under Grant No. 2017YFE0301900, National Science Foundation of China under Grant Nos. 12175053 and 11875233, and ``Users of Excellence program of Hefei Science Center CAS under Contract No. 2021HSC-UE016".
This work has been carried out within the framework of the EUROfusion Consortium, funded by the European Union via the Euratom Research and Training Programme (Grant Agreement No 101052200 - EUROfusion). Views and opinions expressed are however those of the author(s) only and do not necessarily reflect those of the European Union or the European Commission. Neither the European Union nor the European Commission can be held responsible for them.

\section*{References}

\providecommand{\newblock}{}

\end{document}